\documentclass[12pt,epsf]{article}
\usepackage{epsf}
\usepackage{graphicx,graphics,cite,color}
\usepackage{subfigure}
\usepackage[dvipsnames]{xcolor}

\usepackage{graphicx,epsfig}
\usepackage{simplewick}
\usepackage{slashed}
\usepackage{setspace}
\usepackage{hyperref}
\usepackage{empheq}

\usepackage{enumitem}

\newcommand{\ba}{\begin{array}}
\newcommand{\ea}{\end{array}}

\def\lQ{\Lambda_{\rm QCD}}
\newcommand{\nn}{\nonumber}
\newcommand{\be}{\begin{equation}}
\newcommand{\ee}{\end{equation}}
\newcommand{\bea}{\begin{eqnarray}}
\newcommand{\eea}{\end{eqnarray}}
\def\al{\alpha}
\def\als{\alpha_{\rm s}}
\def\siml{{\ \lower-1.2pt\vbox{\hbox{\rlap{$<$}\lower6pt\vbox{\hbox{$\sim$}}}}\ }} 
\def\simg{{\ \lower-1.2pt\vbox{\hbox{\rlap{$>$}\lower6pt\vbox{\hbox{$\sim$}}}}\ }} 

\newcommand{\MS}{\overline{\rm MS}}

\newcommand{\PV}{\rm PV}
\newcommand{\OS}{\rm OS}
\newcommand{\F}{\rm F}

\begin{document}
\title{\vskip-3cm{\baselineskip14pt
}
\vskip1.5cm
The chromomagnetic
moment of a heavy quark with hyperasymptotic precision}
\author{Cesar Ayala$^{1}$ and 
Antonio Pineda$^{2,3}$\\[0.5cm]
{\small ${}^1$  \it Departamento de Ingenier\'ia y Tecnolog\'ias, Sede La Tirana, Universidad de Tarapac\'a, 
}\\
{\small \it Av.~La Tirana 4802, Iquique, Chile} \vspace{0.3cm}\\
{\small ${}^2$ \it Grup de Física Teòrica, Dept. Física,}\\
{\small \it Universitat Autònoma de Barcelona, E-08193 Bellaterra, Barcelona, Spain}\\
{\small ${}^3$ \it Institut de Física d'Altes Energies (IFAE),}
{\small\it The Barcelona Institute of Science and Technology,}\\
{\small\it Campus UAB, 08193 Bellaterra (Barcelona), Spain}
}
\date{}

\maketitle

\thispagestyle{empty}

\begin{abstract}
We determine the normalization of the leading infrared renormalon of the chromomagnetic moment of a heavy quark. Estimates of higher order coefficients of the perturbative series are given. We compute the hyperfine splitting of the $B$ and $D$ mesons for the ground state with hyperasymptotic precision by including the leading terminant, associated with the first infrared renormalon. We fit the experimental data to the operator product expansion theoretical prediction with $\hat \mu^2_{G,\PV}$ as the free parameter. We obtain 
$\hat \mu^2_{G,\PV}=0.507(7)$ GeV$^2$ for the ground state. 
\\[2mm]
\end{abstract}

\newpage
\index
\tableofcontents

\vfill
\newpage

\section{Introduction}
\label{int}

Hadrons containing a single heavy quark can be described efficiently using Heavy Quark Effective Theory (HQET) (for a review, see \cite{Neubert:1993mb}). Its Lagrangian reads
\be
\label{Lag}
{\cal L}=\bar h iD_0 h+\frac{1}{2m_{\OS}}\bar h {\bf D}^2 h +\frac{1}{4m_{\OS}}c_F(\nu)\bar h \sigma_{\mu\nu}gG^{\mu\nu}h+{\cal O}(1/m_{\OS}^2)
\,,
\ee
where $m_{\OS}$ is the on-shell mass.

In this paper, we focus on the hyperfine splitting of the $B$ and $D$ mesons. They can be written in terms of an operator product expansion-like expansion in powers of $1/m$ \cite{Bigi:1994ga}:
\be
\label{HF}
M_V-M_P=\frac{2}{3}\frac{c_F(\nu)}{m_{\OS}}\mu_G^2(\nu)
+\frac{1}{3m_{\OS}^2}
\left[
c_F(\nu)\rho_{\pi G}^3(\nu)+c_F^2(\nu)\rho_A^3(\nu)-(2c_F(\nu)-1)\rho^3_{LS}(\nu)
\right]
\ee
where in the last term reparameterization invariance \cite{Luke:1992cs} is used to relate the Wilson coefficient $c_{LS}$ with $c_F$. 

We also have that (the heavy meson is at rest)
\be
\mu_G^2=-\frac{1}{2}\langle B|\bar h \sigma_{\mu\nu}gG^{\mu\nu}h |B \rangle
\ee
with the states normalized in a nonrelativistic way: $\langle B({\bf p}')|B({\bf p}) \rangle=(2\pi)^3\delta^{(3)}({\bf p}'-{\bf p})$.

$\mu_G^2$ is heavy quark mass independent. Therefore, it is the same for $B$ and $D$ mesons. On the other hand it does depend on the quantum number: it is different for the ground state and for each of the $B/D$ meson excitations. The other nonperturbative matrix elements follow from subleading effects in the $1/m$ expansion of the HQET Lagrangian.

The leading term in this $1/m$ expansion is proportional  to the chromomagnetic Wilson coefficient: $c_F$. This term has been computed to order $\alpha_s^3$ \cite{Amoros:1997rx,Czarnecki:1997dz,Grozin:2007fh}, as well as its anomalous dimension. The perturbative expression of $c_F(\nu)$ to order $\alpha_s^3$ can be found in Eq. (14) of \cite{Grozin:2007fh}. Traditionally, $c_F(\nu)$ is written in a theory with four active flavours for the $B$ meson case. We display it as follows 
\be
 c_F(\nu)=1+\sum_{n=0}^{\infty}C^{(+)}_n\alpha_{s,(+)}^{n+1}(m_{\OS})\,.
\ee
The label + stands for a theory with $n_l+1$ active quarks with $n_l$(=3) light quarks and one massive (charm) quark.

Mass charm effects start to appear at ${\cal O}(\als^2)$. Therefore, we split the associated coefficient in the following way: 
\be
C_1^{(+)}=C_1^{(+)}(m_{c,\OS}=0)+\delta C_1^{\rm charm}
\,.
\ee

$\delta C_1^{\rm charm}$ was computed in Ref. \cite{Davydychev:1998si}:
\be
\delta C_1^{\rm charm}=\frac{T_F}{(4\pi)^2}\left(8C_FA_F+\frac{4}{3}C_AA_A\right)
\,,
\ee
where ($r=m_{c,\OS}/m_{b,\OS}$)
\be
A_F=-r(1+r)(1-r-4r^2)L_++r(1-r)(1+r-4r^2)L_-+6r^2\left(\ln(r)+\frac{4}{3}\right)\,,
\ee
\be
A_A=-(1+r)(2+4r-r^2)L_+-(1-r)(2-4r-r^2)L_-+2\ln^2(r)+\frac{\pi^2}{3}+2r^2(\ln(r)+1)\,,
\ee
and
\be
L_+= {\rm Li}_2(-r^{-1})+\ln(1/r)\ln(1+1/r) \qquad
L_-=-{\rm Li}_2(1-1/r)+\frac{\pi^2}{6}
\,.
\ee
The above expression applies to $B$ mesons considering that one has four active quarks ($n_l=4$), with the qualification that it includes the leading finite charm quark mass effects. For the $D$ mesons one has three active flavours ($n_l=3$) and sets $\delta C_1^{\rm charm}$ to zero.

\subsection{Charm quark decoupling}

The size of the correction associated to the finite charm quark mass for $\bar m_{b}=4.186$ GeV and $\bar m_{c}=1.223$ GeV, which we take from Ref. \cite{Peset:2018ria} (with this precision the difference between the pole and $\MS$ mass is negligible), reads
\be
\label{nodec}
\delta C_1^{\rm charm}\alpha_{s,(+)}^2(\bar m_{b})=0.006
\,,
\ee
where we recall that the strong coupling constant has four active flavours. This correction is quite small in comparison with the size of the ${\cal O}(\als^2)$ term without finite mass effect:
\be
 C_1^{(+)}(m_{c,\OS}=0)\alpha_{s,(+)}^2(\bar m_{b})=0.087
 \,.
\ee
Nevertheless, this last number can be contaminated by renormalons, so the real size could be smaller. On top of that, there is another issue that should be considered concerning charm quark effects. As we have already mentioned, we apply this formalism to the hyperfine splitting of the $B$ and $D$ mesons. For the latter, it is clear that the number of active fermions is $n_l=3$. For the $B$ meson, one may think, a priori, that one should take $n_l=4$  instead, by considering the charm as a light quark, since the typical scale in the problem is the heavy quark mass, $m_b$. This is how the number in Eq. (\ref{nodec}) has been obtained. Nevertheless, as it has been argued in \cite{Ayala:2014yxa}, as we go to high orders in perturbation theory, the typical scale in the loop is not $m_b$ but rather $m_be^{-n}$, where $n$ is the loop order. Therefore, it is more convenient to work with $n_l=3$ also in the $B$ meson case. This has proven to yield a more convergent perturbative series in the case of heavy quarkonium for the $\Upsilon$ case. This is the attitude that we will also take in this paper. Therefore, we have to decouple the charm quark from the strong coupling constant. After decoupling $\delta C_1^{\rm charm}  \longrightarrow \delta C_{1,\rm dec.}^{\rm charm}$ where
\be
\delta C_{1,\rm dec.}^{\rm charm}=\delta C_{1}^{\rm charm}+[C^{(+)}_1(m_c=0)-C_1(m_c=0)]+C_0\frac{T_F}{3\pi}\ln\left(\frac{\bar m_b^2}{\bar m_c^2}\right)
\,,
\ee
with $C_1(m_c=0)=C_1(n_l=3;m_c=0)$. 

Let us check that this procedure leads to much smaller charm quark effects (note that now $\als$ runs with three active light quarks). We have 
\be
\delta C_{1,\rm dec.}^{\rm charm}\als^2(\bar m_b)=-0.0003
\,.
\ee
We see that the size of this correction is smaller than the analogous one without decoupling by an order of magnitude, confirming the cancellation. 

\subsection{General formulas}

In the following, we will always work with three active flavours and, in the case of the bottom, we add $\delta C_{1,\rm dec.}^{\rm charm}\als^2(\nu_h)$ (with $\nu_h \sim m_{\OS}$) to the initial matching condition. In any case, its effect will be very small.

We now proceed to implement the resummation of large logarithms. 
The renormalization group improved expression of $c_F$ reads (adapting the notation of \cite{Grozin:1997ih} to ours)
\be
c_F(\nu)=c_F(\nu_h)exp\left(-\int_{\als(\nu)}^{\als(\nu_h)}d\als\frac{\gamma(\als)}{\beta(\als)}
\right)
\equiv \hat c_F K(\nu)
\,,
\ee
where ($\beta_0=11/3C_A-4/3T_fn_l$,...)
\be
\nu\frac{d \als}{d\nu}=\beta(\als)=-2\als\left(\beta_0\frac{\als}{4\pi}+\beta_1\left(\frac{\als}{4\pi}\right)^2+\cdots \right)
\,,
\ee
and
\be
\gamma(\als)=\gamma_0\frac{\als}{4\pi}+\gamma_1\left(\frac{\als}{4\pi}\right)^2+\cdots
\ee
with $\gamma_0=2C_A$, $\gamma_1=2C_A(13\beta_0-25C_A)$, .... By default we will take $\nu_h= m_{\OS}$. 

The perturbative expression of $\hat c_F$ reads ($\gamma_{\F}\equiv\gamma_0/(2\beta_0)$)
\be
\hat c_F=\left[\als(m_{\OS})\right]^{\gamma_{\F}}\left[1+c(m_{\OS})\right]\;,\qquad 
c(m_{\OS})=\sum_{n=0}^{\infty}c_n\als^{n+1}(m_{\OS})\,.
\ee

It is convenient to reexpress Eq. (\ref{HF}) in terms of renormalization group invariant quantities (within perturbation theory at least). We then have that the hyperfine splitting reads \cite{Grozin:1997ih}
\be
\label{HFhat}
M_V-M_P=\frac{2}{3}\frac{\hat c_F}{m_{\OS}}\hat \mu_G^2
+\frac{1}{3m_{\OS}^2}
\left[
\hat c_F \hat \rho_{\pi G}^3+{\hat c}_F^2 {\hat \rho}_A^3-(2\hat c_F-1)\hat \rho^3_{LS}
\right]
\,.
\ee

In Eq. (\ref{HFhat}) all its constituents are renormalization-scale and scheme independent to any finite order in perturbation theory. This is not enough, we want exponential precision: $\sim e^{-1/\als}$ (or power-like, $\Lambda_{\rm QCD}/m$, precision in terms of $\Lambda_{\rm QCD}$). To reach such precision, we have to overcome the problem that perturbative expansions are asymptotic. Therefore, each term of the OPE expression written in Eq. (\ref{HFhat}) is ill-defined. Thus,  Eq. (\ref{HFhat}) is a formal expression, and, without further qualifications, it is of little use. The origin of this problem comes because, even though the hyperfine splitting is an observable, and, therefore, well defined, the splitting between the different terms of the OPE is ambiguous. When working in dimensional regularization using minimal-like subtraction schemes, this reflects in the fact that the perturbative series of $c(m_{\OS})$ is asymptotically divergent. Therefore, its sum does not converge to a number, and a method has to be used to regularize the perturbative sum. A suitable one is first to construct the Borel transform of the perturbative sum and, afterwards, to do the inverse of the Borel transform (also named Borel sum or integral). Such an inverse is still ill-defined due to singularities on the positive axis of the Borel plane. The location and 
character of such singularities are determined by the OPE \cite{Parisi:1978bj,Mueller:1993pa}. This information is enough to determine the divergence pattern of the perturbative series up to an overall normalization, which is not fixed by the OPE. Such a divergent pattern of the perturbative expansion associated with the OPE is usually referred to as renormalons \cite{tHooft}.

To fully fix the leading term of the OPE with exponential accuracy, we have to specify how we handle the singularities in the Borel plane of $c(m_{\OS})$ when doing the Borel sum (we assume that there are no renormalons in the anomalous dimension). Consequently, $c(m_{\OS})$ is going to be our main subject of study. Following the discussion in \cite{Ayala:2019uaw}, we use the Principal Value (PV) prescription for the Borel integral (the median resummation above and below the real axis). The reason is that the outcome is expected to be real and scale/scheme independent (see the discussion in Refs. \cite{Ayala:2019uaw,Takaura:2020byt}). 

Defining $\hat c_F$ with exponential accuracy also defines the different power corrections. Therefore, we can rewrite Eq. (\ref{HFhat}) as
\be
\label{HFPV}
M_V-M_P=\frac{2}{3}\frac{\hat c_{F,\PV}}{m_{\PV}}\hat \mu_{G,\PV}^2
+\frac{1}{3m_{\PV}^2}
\left[
\hat c_{F,\PV} \hat \rho_{\pi G,\PV}^3+{\hat c}_{F,\PV}^2 {\hat \rho}_{A,\PV}^3-(2\hat c_{F,\PV}-1)\hat \rho^3_{LS,\PV}
\right]
\,,
\ee
where (often we will work with the variable $u=\frac{\beta_0 t}{4\pi}$ instead of $t$)
\bea
\label{cPV}
&&
\hat c_{F,\PV}=\left[\als(m_{\PV})\right]^{\gamma_{\F}}[1+c_{\PV}(m_{\PV})] 
\\
\nn
&&
 c_{\PV}(m_{\PV})=\int_{0,\rm PV}^{\infty} dt\; e^{-t/\als(m_{\PV})}B[c](t)\,, \qquad B[c](t)=\sum_{n=0}^{\infty}
\frac{c^{(s)}}{s!}t^s
\,,
\eea
\be
\label{mPV}
m_{\PV}=\bar m+\int_{0,\rm PV}^{\infty} dt\; e^{-t/\als(m_{\PV})}B[m_{\PV}](t)\,, \qquad B[m_{\PV}](t)= \sum_{n=0}^{\infty}
\frac{r_n}{n!}t^n
\,,
\ee
and
\be
\hat \mu_{G,\PV}^2 \equiv  \mu_{G,\PV}^2(\nu)\left[\als(\nu)\right]^{-\gamma_{\F}}
\,.
\ee
We have absorbed the prefactor $\left[\als(\nu)\right]^{-\gamma_{\F}}$ in  
$\hat \mu_{G,\PV}^2$. It is usually stated that, after introducing such a prefactor, $\hat \mu_{G,}^2$ is renormalization group invariant. We would like to emphasize that this is not necessarily so. In order $\hat \mu_{G,\PV}^2$ to be renormalization scale and scheme independent, it is necessary that the regularization of the perturbative sum of 
$c_{\PV}(m_{\PV})$ and $m_{\PV}$ is made in such a way that it is explicitly scale and scheme independent (something that the PV summation scheme delivers). Note also that all nonperturbative constants are defined in a theory with three active light quarks. Therefore, they are the same for $B$ and $D$ physics. Finally, since we work with $\hat c_F$, the four-loop anomalous dimension of $\gamma_3$ enters in $c_2$. At present, such four-loop anomalous dimension is unknown. Therefore, we will set it to zero. This should not significantly affect our renormalon-based determinations, as such a coefficient should be renormalon-independent (or, at least, subleading compared with those considered in this paper). We have performed a couple of tests to check this hypothesis: i) We have seen how the renormalon behaves if working directly with $c_F$, for which the three terms of the perturbative expansion are known exactly, or ii) by adding the large $\beta_0$ estimate of $\gamma_3$ to our analyses. 

Assuming the validity of the OPE in its nonperturbative version \cite{Shifman:1978bx} (which we take for granted), $\hat \mu_{G,\PV}^2$ can be written as $\hat \mu_{G,\PV}^2=\bar \mu_{G,\PV}^{2,\MS} \Lambda_{\MS}^2$, where $\bar \mu_{G,\PV}^{2,\MS}$ is a dimensionless constant. Let us emphasize that such equality holds irrespectively of the scheme used for the strong coupling constant. Therefore, we can also generically write 
\be
\label{eq:hatf3PV}
\hat \mu_{G,\PV}^{2}=\bar \mu_{G,\PV}^{2,X} \Lambda_{\rm X}^2=\bar \mu_{G,\PV}^{2,X}
\left(\frac{\beta_0\alpha_X(\nu)}{4\pi}\right)^{-2b}e^{-\frac{4\pi}{\beta_0\alpha_X(\nu)}}\left(1+{\cal O}(\alpha_X)\right)
\,,
\ee
 where X stands for the renormalization scheme of the strong coupling constant and 
we define $b={\beta_1/(2\beta_0^2)}$ (where $b=51/121\simeq 0.421$ for $n_l=0$ and $b=32/81\simeq 0.395$ for $n_l=3$). Therefore, provided $\bar \mu_{G,\PV}^{2,X}$ is obtained in one scheme, one can easily transform it to a different scheme using the very same conversion factor one uses to transform $\lQ$ from one scheme to another. In the present work $X=\MS$.

Whereas the above procedure yields unambiguous and convenient definitions of the different terms of the OPE, this does not mean that we have the means to compute them. In practice, we only have a relatively small set of the first order terms of the perturbative expansion, and the experimental (or lattice) data. Nowadays, it is not possible to compute the nonperturbative corrections from first principles via analytic methods. It is only possible to compute them in some cases, numerically, from lattice simulations. Nevertheless, even such computations are  unavoidably plagued by perturbative corrections. Actually, those are the dominant contributions to the observable. A paradigmatic case is the computation of the gluon condensate in the lattice, where its value is orders of magnitude bigger than the actual size of the nonperturbative gluon condensate \cite{Bali:2014sja,Ayala:2020pxq}. Overall, to fit the nonperturbative condensate to lattice or experimental data, it is unavoidable to compute the perturbative series with exponential (in $-1/\al(m_{\PV})$) accuracy, or with power (in $1/m_{\PV}$) accuracy, first. 

Whereas it is not possible to obtain the exact expression of $c_{\rm PV}$, it can be computed approximately, and, most importantly, with a well-defined method to quantify the error in a parametric way \cite{Ayala:2019uaw,Ayala:2019lak,Ayala:2019hkn}. This method adapts the hyperasymptotic expansion used in ordinary differential equations \cite{BerryandHowls} (see also \cite{Dingle}) to the case of quantum field theories with marginal operators, and it has successfully been applied to a variety of observables \cite{Ayala:2019hkn,Ayala:2020odx,Ayala:2020pxq}.  Therefore, it is our plan to apply such a method to $c_{\rm PV}(m_{\PV})$. We will see later that the perturbative series of $c_{\rm PV}(m_{\PV})$ is known to high enough orders to start showing its asymptotic nature for the values of the bottom and charm quark masses. Therefore, this opens the venue to determine the hyperfine splitting of the $B$ and $D$ mesons with exponential accuracy. This is very interesting, since it will allow us to determine the leading nonperturbative correction, $\hat \mu_{G,\PV}^{2}$, with hyperasymptotic precision. This object plays an important role in $B$-physics determinations of some CKM matrix elements. We plan to address this issue in the future. 

Overall, we have now turned the problem into evaluating $c_{\PV}$ with the highest possible accuracy, i.e. including power corrections. We address this goal in Sec. \ref{Sec:cPV}.  

The paper is organized as follows. 
In Sec. 2, $c_{\PV}$ will be computed with hyperasymptotic precision, and new estimates of the higher order coefficients of the perturbative series, are given. 
In Sec. 3, the comparison with the experimental data will be made allowing us the extraction of $\hat \mu_{G,\PV}^2$. Finally, 
the conclusions are presented in Sec. 4.

\section{Hyperasymptotic approximation to $c_{\PV}$}
\label{Sec:cPV}

\subsection{Renormalons}
We first need to know the renormalon structure of the perturbative series. We take the results relevant to our case from the analysis made in Ref. \cite{Grozin:1997ih}.

The Borel transform near the closest infrared renormalon singularity has the following structure \cite{Grozin:1997ih}:
\bea
\label{SX}
B[c](t(u))&=& {\nu \over m_{\PV}}
\left(
Z_m+
 \frac{1}{2}(Z_{\pi G}-2Z_{LS})
\right)
\frac{1}{(1-2u)^{1+b}}(1+{\cal O}(1-2u)) 
\nn
\\
&&
+
 \frac{1}{2}
{\nu \over m_{\PV}}Z_A
\frac{1}{(1-2u)^{1+b-\gamma_{\F}}}(1+{\cal O}(1-2u)) 
\nn
\\
&&
+
 \frac{1}{2}
{\nu \over m_{\PV}}Z_{LS}
\frac{1}{(1-2u)^{1+b+\gamma_{\F}}}(1+{\cal O}(1-2u)) 
+c_{reg}(u)
\,,
\eea
where $c_{reg}(u)$ is an analytic function at $u=1/2$. 

In all cases, the ${\cal O}(1-2u)$ corrections can be determined from the perturbative expansions of $\Lambda_{\MS}$, $m_{\PV}$ and $c(m_{\PV})$. For instance, the ${\cal O}(1-2u)$ corrections of the term proportional to $Z_m$ follows from the ratio of the perturbative expansion of $c(m_{\PV})$ and $m_{\PV}$. Nevertheless, the effect of these corrections gets masked by our lack of knowledge of the normalization of the different renormalons. Therefore, we do not consider them in detail. 

With the above precision for the Borel transform, we can determine the large $n$ behavior of the $c_n$ coefficients:
\bea
\label{CXRG}
&&
c_n
\stackrel{n\rightarrow\infty}{=} 
\left(Z_m+\frac{1}{2}(Z_{\pi G}-2Z_{LS})\right)\,{\nu \over m_{\PV}}\,
\left({\beta_0 \over 2\pi}\right)^n
\,{\Gamma(1+b+n) \over
\Gamma(1+b)}
\\
&&
\nn
+
\frac{1}{2}Z_A\,{\nu \over m_{\PV}}\,
\left({\beta_0 \over 2\pi}\right)^n
\,{\Gamma(1-\gamma_{\F}+b+n) \over
\Gamma(1+b-\gamma_{\F})}
{}_1F_1\left(\gamma_{\F},-b+\gamma_{\F}-n,\ln(m_{\PV}^2/\nu^2)\right) 
\\
&&
\nn
+
\frac{1}{2}Z_{LS}\,{\nu \over m_{\PV}}\,
\left({\beta_0 \over 2\pi}\right)^n
\,{\Gamma(1+\gamma_{\F}+b+n) \over
\Gamma(1+b+\gamma_{\F})}
{}_1F_1\left(-\gamma_{\F},-b-\gamma_{\F}-n,\ln(m_{\PV}^2/\nu^2)\right) 
.
\eea
The above expression contains subleading terms in the $1/n$ expansion. 
In the strict $1/n$ expansion, 
it simplifies to:
\bea
\label{CXRG1overn}
&&
c_n
\stackrel{n\rightarrow\infty}{=} 
\left(Z_m+\frac{1}{2}(Z_{\pi G}-2Z_{LS})\right)\,{\nu \over m_{\PV}}\,
\left({\beta_0 \over 2\pi}\right)^n
\,{n!n^b \over
\Gamma(1+b)}
\\
&&
\nn
+
\frac{1}{2}Z_A\,{\nu \over m_{\PV}}\,
\left({\beta_0 \over 2\pi}\right)^n
\,{n!n^bn^{-\gamma_{\F}} \over
\Gamma(1+b-\gamma_{\F})}
\left(1+{1 \over n}\ln(m_{\PV}^2/\nu^2)\right)^{-\gamma_{\F}} 
\\
&&
\nn
+
\frac{1}{2}Z_{LS}\,{\nu \over m_{\PV}}\,
\left({\beta_0 \over 2\pi}\right)^n
\,{n!n^bn^{\gamma_{\F}} \over
\Gamma(1+b+\gamma_{\F})}
\left(1+{1 \over n}\ln(m_{\PV}^2/\nu^2)\right)^{\gamma_{\F}} 
.
\eea

In Eq. (\ref{SX}), the last term produces the strongest singularity in the Borel plane. Therefore, we expect we can determine $Z_{LS}$ by neglecting the first and second line of Eq. (\ref{SX}).  The formal precision is of order 
$\als^{\gamma_{\F}} \sim \als^{1/3}$ for $n_l=3$ and $\als^{3/11}\sim \als^{0.28}$ for $n_l=0$. Within this approximation, one can consider different alternatives. One is to work with the quantity (something similar was made in Ref. \cite{Ayala:2022mgz}) 
\be
\label{tildeCB}
\tilde  c(m) \equiv c(m)\left[\frac{\als(m)}{\als(\nu)}\right]^{\gamma_{\F}}=\sum_{n=0}^{\infty}\tilde c_n\als^{n+1}
\,
\ee
and focus in the leading renormalon. Then Eq. (\ref{tildeCB}) takes the form
\be
\label{tildeCXRG}
\tilde c_n
\stackrel{n\rightarrow\infty}{=}
\frac{1}{2}Z_{LS}\,{\nu \over \bar m}\,
\left({\beta_0 \over 2\pi}\right)^n
\,{\Gamma(1+\gamma_{\F}+b+n) \over
\Gamma(1+b+\gamma_{\F})} 
\left(1+{\cal O}\left((1/n^{\gamma_{\F}}\right)\right)
\,,
\ee
or
\be
\tilde c_n
\stackrel{n\rightarrow\infty}{=}
\label{tildeCXRG1overn}
\frac{1}{2}Z_{LS}\,{\nu \over \bar m}\,
\left({\beta_0 \over 2\pi}\right)^n
\,{n!n^bn^{\gamma_{\F}} \over
\Gamma(1+b+\gamma_{\F})}
\left(1+{\cal O}\left((1/n^{\gamma_{\F}}\right)\right)
\,.
\ee

\subsection{Determination of the normalization constant}
\label{Sec:Norm}

\begin{figure}
\begin{center}
\includegraphics[width=.5\textwidth]{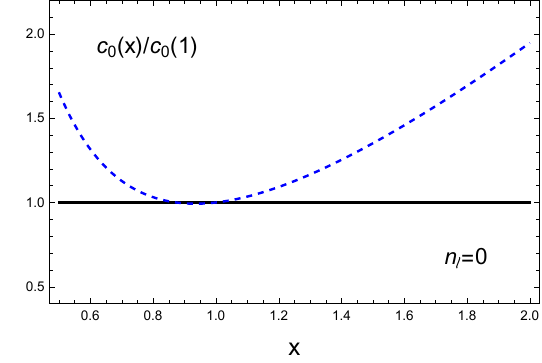}
\includegraphics[width=.5\textwidth]{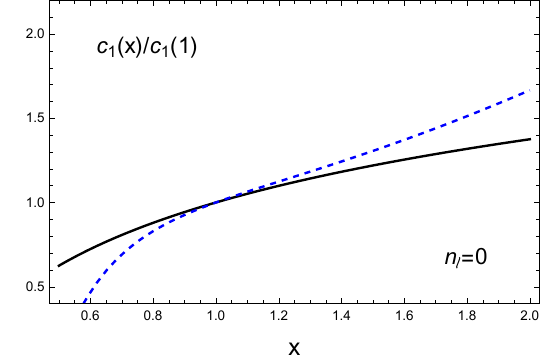}
\includegraphics[width=.5\textwidth]{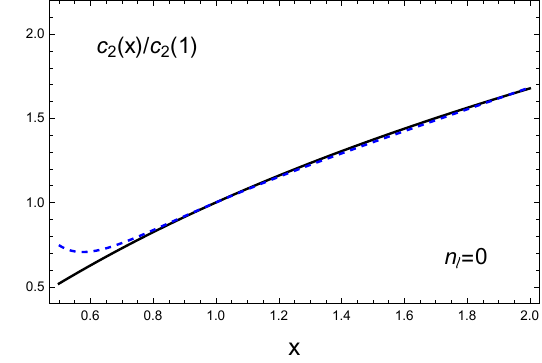}
\end{center}
\caption{Plot of the scale dependence of the ratio of $c_n(x)/c_n(1)$, where $x=\nu/m_{\PV}$, $n_l=0$, and $n=0$, 1 and 2. The black solid line corresponds to using the exact expression for the perturbative coefficients. The blue dashed line corresponds to using the asymptotic expression for the perturbative coefficients.}
\label{Fig:ratiocnnf0}
\end{figure}

\begin{figure}
\begin{center}
\includegraphics[width=.5\textwidth]{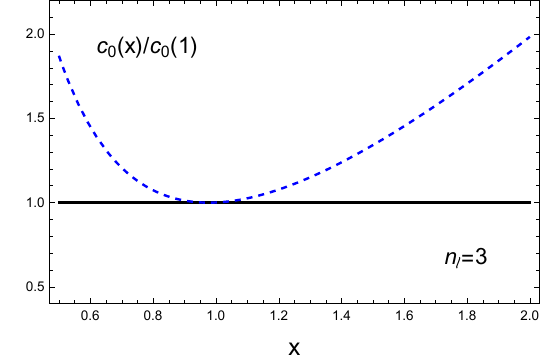}
\includegraphics[width=.5\textwidth]{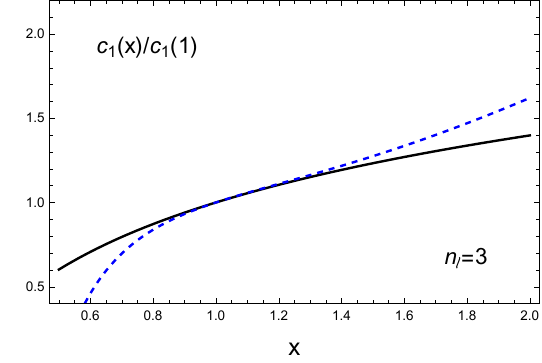}
\includegraphics[width=.5\textwidth]{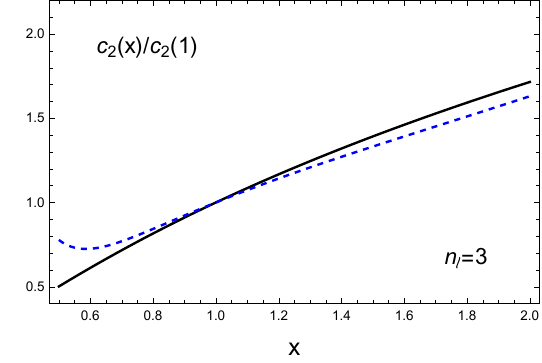}
\end{center}
\caption{Plot of the scale dependence of the ratio of $c_n(x)/c_n(1)$, where $x=\nu/m_{\PV}$, $n_l=3$, and $n=0$, 1 and 2. The black solid line corresponds to using the exact expression for the perturbative coefficients. The blue dashed line corresponds to using the asymptotic expression for the perturbative coefficients.}
\label{Fig:ratiocnnf3}
\end{figure}

In this subsection, we obtain the normalization of the leading infrared renormalon: $Z_{LS}$. Before we proceed, we want to check that the perturbative coefficients have reached their asymptotic regime. A very visual way to see this is by confirming that the logarithmic dependence in the renormalization scale of the perturbative coefficients effectively become linear as we increase $n$. We indeed see this behavior, particularly for $n_l=0$, but also for $n_l=3$, in Figs. \ref{Fig:ratiocnnf0} and \ref{Fig:ratiocnnf3}, respectively.\footnote{The relative weakness of the renormalon signal as we increase the number of light quarks is to be expected on general grounds. Typically, for $n_l \sim 6$ the renormalon signal becomes quite weak (see for instance the discussion in Ref. \cite{Ayala:2014yxa}).}

We now proceed with the actual determination of $Z_{LS}$. 
It has been observed in Refs. \cite{Bali:2013pla,Ayala:2014yxa} that a more stable result is obtained by determining the normalization directly from the ratio of the exact and asymptotic expression. Therefore, this is the path we will follow in this paper. We will use Eq. (\ref{CXRG}) for the asymptotic expression and set the first and second line to zero, as they are subleading by powers of $1/n^{\gamma_F}$. Therefore, we have 
\bea
\nn
&&
c_n^{(\rm as)}
\stackrel{n\rightarrow\infty}{\simeq} 
\frac{Z_{LS}}{2}\,{\nu \over m_{\PV}}\,
\left({\beta_0 \over 2\pi}\right)^n
\,{\Gamma(1+\gamma_{\F}+b+n) \over
\Gamma(1+b+\gamma_{\F})}
{}_1F_1\left(-\gamma_{\F},-b-\gamma_{\F}-n,\ln(m_{\PV}^2/\nu^2)\right) 
.
\\
\label{CXRGZLS}
\eea
and determine $Z_{LS}$ using the following equality ($x=\nu/m_{\PV}$):
\be
\label{ZLSdef}
Z_{LS}(x)=Z_{LS}\frac{c_n(x)}{c_n^{(\rm as)}(x)}
\,.
\ee

To determine the central value of our determination, we use this expression with $n=2$ and $x=1$. 
To check how robust this determination is, we perform a series of tests, which allow us to determine the error.
The error in our determination of the normalization of the infrared renormalon is due to our incomplete knowledge of the perturbative series and of the OPE. This reflects that the result will depend on the scale, the order in which we truncate, and the explicit expression we use to determine the normalization. We use these three methods as indicators of the error:
\begin{enumerate}[label=(\roman*)]
\item
We determine the variation of $Z_{LS}$ if we change $x$ in the range $(1/\sqrt{2},\sqrt{2})$.
\item
We determine the difference with the determination of  $Z_{LS}$ if we set $n=1$ instead of $n=2$. 
\item
The other renormalons located at $u=1$ can significantly affect the determination of $Z_{LS}$, as they are suppressed by a relatively small factor, $1/n^{\gamma_F}$, due to the different anomalous dimension of these renormalons. To account for this, we perform a fit of Eq. (\ref{ZLSdef}) using Eq. (\ref{CXRG}) in the range $x=(1/\sqrt{2},\sqrt{2})$, allowing also $Z_A$, $Z_{\pi G}$ and $Z_m$ to be different from zero. In this fit we fix $Z_m$ to the value 
obtained in Ref. \cite{Ayala:2025trr} (note that $Z_m$ always appears in a fixed combination with $Z_{\pi G}$. Therefore, it would be impossible to fit them separately). The outcome reads 
\bea
&&
\label{fitnf0}
n_l=0 \quad Z^{\rm fit}_{\pi G}=1.744\,, \quad Z^{\rm fit}_{LS}=1.764\,, \quad Z^{\rm fit}_A= 0.4603 \,,
\\
&&
\label{fitnf3}
n_l=3 \quad Z^{\rm fit}_{\pi G}=1.148\,, \quad Z^{\rm fit}_{LS}=0.872\,, \quad Z^{\rm fit}_A= 0.1175
\,.
\eea
The value of $Z_{LS}$ is relatively close to our central value. Still, we find this produces the largest error in the determination of $Z_{LS}$ (the difference with the central value reads 0.18
for $n_l=0$ and -0.42
for $n_l=3$), particularly for $n_l=3$. This indicates that there is a sizeable mixing with the other renormalons. 
\end{enumerate}

The three methods yield different ways to measure the fact that $n$ is not infinity. We combine them in quadrature, but the error is overwhelmingly saturated  by the difference with the "fit" value. We show our results in Table \ref{tableZLS}.
\begin{table}[htb!]
\begin{center}
\begin{tabular}{|c|c|c|}
\hline
$n_l$&$0$ &$3$\\\hline
$Z_{LS}$&$1.58(19)$&$1.29(44)$\\ \hline
\end{tabular}
\end{center}
\caption{\it Values of $Z_{LS}$ for $n_l=0$ and $n_l=3$.}
\label{tableZLS}
\end{table}

\begin{figure}
\begin{center}
\includegraphics[width=.7\textwidth]{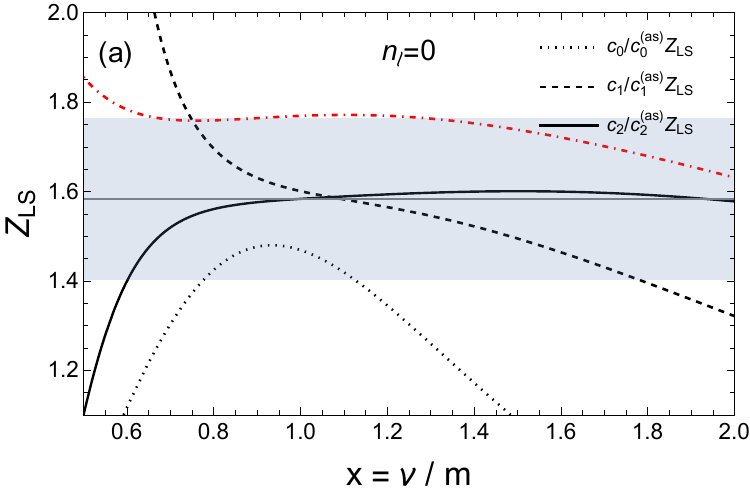}
\includegraphics[width=.7\textwidth]{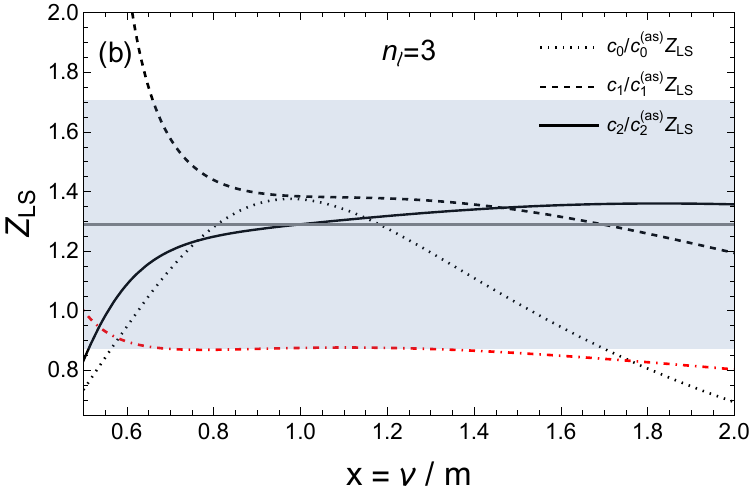}
\end{center}
\caption{ Determination of $Z_{LS}$ for $n_l=0$ and $n_l=3$. The dotted, dashed and solid black lines correspond to Eq. (\ref{ZLSdef}) for $n=0$, 1 and 2 respectively, where $x=\nu/m_{\PV}$. The dash-dotted red line corresponds to $Z^{\rm fit}_{LS}\frac{c_n(x)}{c_{n,\rm fit}^{(\rm as)}(x)}$ using the normalizations obtained in Eqs. (\ref{fitnf0}) and (\ref{fitnf3}). The horizontal line and the blue band yield our prediction for $Z_{LS}$ and the error estimate listed in Table \ref{tableZLS}.}
\label{Fig:ZLS}
\end{figure}

The above discussion is summarized in Fig. \ref{Fig:ZLS}. In these figures we plot $Z_{LS}\frac{c_n(x)}{c_n^{(\rm as)}(x)}$ for $n=0$, 1, 2 in the range $(1/\sqrt{2},\sqrt{2})$ forn $n_l=0$ and $n_l=3$. We can see that the scale dependence becomes smother as we go to higher orders (both for $n_l=0$ and $n_l=3$). We also plot $Z^{\rm fit}_{LS}\frac{c_n(x)}{c_{n,\rm fit}^{(\rm as)}(x)}$ using the normalizations obtained in Eqs. (\ref{fitnf0}) and (\ref{fitnf3}). Note that the difference with our central value listed in Table \ref{tableZLS}, $Z_{LS}$, does not exactly corresponds to $Z^{\rm fit}_{LS}-Z_{LS}$ but it gives a good measure of the scale dependence of $c_n^{(as)}$ obtained from the fit. Finally, in Fig. \ref{Fig:ZLS}, we also plot the central value and the error band obtained in Table \ref{tableZLS}.
 
We have also considered other possibilities to estimate the error. They yield results well inside errors.  One possible modification is to work with the heavy quark $\MS$ mass instead of working with the pole mass. This introduces an extra renormalization scale dependence not associated with renormalons. Consequently, the scale dependence of the result increases but it is well inside errors. Another possibility is to eliminate the anomalous dimension of the leading infrared renormalon, i.e. to work with Eq. (\ref{tildeCB}) and (\ref{tildeCXRG}). Again, the differences are small (actually, such an approach produces slightly more stable results).

\medskip

\noindent
{\bf Large $\beta_0$}\\
$Z_{LS}$ has been computed before in the large $\beta_0$ approximation \cite{Grozin:1997ih}. In our conventions we get (in the $\MS$ scheme)
\be
\frac{Z_{LS}}{2}=\frac{e^{5/6}}{4\pi}C_A 2^{b+\gamma_{\F}}K_4\Bigg|_{\beta_0}=0.549318
\,,
\ee
where in the last equality we have set in the large $\beta_0$ limit by setting 
$K_4=1$, and $b=\gamma_{\F}=0$. This number is not very much off the numbers we obtain in Table \ref{tableZLS} (note that our number depends on $n_l$). On the other hand, the large $\beta_0$ approximation predicts the following equalities
\be
\label{largeb0}
Z_m^{(\beta_0)}-\frac{Z^{(\beta_0)}_{LS}}{2}+\frac{Z^{(\beta_0)}_{\pi G}}{2}+\frac{Z^{(\beta_0)}_{A}}{2}
=
2.90
\,,
\ee
and
\be
-\frac{Z^{(\beta_0)}_{LS}}{2}+\frac{Z^{(\beta_0)}_{\pi G}}{2}+\frac{Z^{(\beta_0)}_{A}}{2}
=
1.92
\,,
\ee
which do not compare well with the numbers obtained in Eqs. (\ref{fitnf0}) and (\ref{fitnf3}). This is not a problem since the large $\beta_0$ limit is not necessarily a good approximation for the determination of the normalization, besides having some degree of ambiguity (see the discussion in Ref. \cite{Bali:2013pla}). We recall that also in the case of the pole mass, the large $\beta_0$ result was off by around a factor of two. 

On the other hand we can test the large $\beta_0$ limit by setting $n_l \rightarrow \infty$. We observe a tendency to approach the value obtained in Eq. (\ref{largeb0}) but still with significant differences. If we work in the large $\beta_0$ approximation (where we know all terms of the perturbative expansion), renormalon dominance needs of higher order coefficients than those we have in this paper to get accurate determinations of the normalization of the renormalon.

\medskip

\noindent
$\boldsymbol{c_n^{(as)}}$

We are now in the position to give predictions for the high-order coefficients of the perturbative series. We displayed them for the cases of $n_l=0$ and $n_l=3$ in Table \ref{tablecn}. We should stress that our numbers incorporate the right asymptotic behavior, which is not the case for large-$\beta_0$ estimates. The central value of $c_n^{(as)}$ is determined using Eq. (\ref{CXRGZLS}) together with the values of $Z_{LS}$ obtained in Table \ref{tableZLS}. The determination of the error of the coefficients is made by considering the same sources of error we used for $Z_{LS}$ and combining them in quadrature:
(i) We change $x$ in the range $(1/\sqrt{2},\sqrt{2})$; 
(ii) We determine the difference with the determination of  $c_n^{(as)}$ if we set $n=1$ instead of $n=2$;
(iii) We determine the difference with the determination of  $c_n^{(as)}$ using Eq. (\ref{CXRG}) with the values of $Z_A$, 
$Z_{\pi}$ and $Z_{LS}$ obtained in Eqs. (\ref{fitnf0}) and (\ref{fitnf3}).

It is interesting to see the differences with the error of $Z_{LS}$. For $Z_{LS}$ the error was completely dominated by item (iii), and particularly large for $n_l=3$. For $c_n^{(as)}$ the relative size between the different sources of error is more equilibrated. For $n_l=0$ the error is dominated by item (i) (except for $n=0$). This trend gets magnified as we increase $n$. For $n_l=3$ we find that the three sources of error are of similar size. Overall, we find that we can determine the asymptotic behavior of the perturbative coefficients more accurately than $Z_{LS}$. This is because the determination of the individual normalizations of the renormalons can be affected by large errors. This is due to the fact that these renormalons are located at the same point in the Borel plane and that their difference only comes from the strength of the singularity. This makes that the $x$ dependence of the different renormalons in the perturbative coefficients is similar. This is the reason we have only attempted to determine the normalization of the renormalon with the strongest singularity. A more thorough analysis would then be necessary to determine the individual normalizations, as they are highly correlated. For $c_n$ such an analysis is not necessary, as we are only interested in the combinations that appear in the coefficient. In any case, what it still holds is that the relative error is bigger for $n_l=3$ compared with the $n_l=0$ case (but of similar absolute size). 

\begin{table}[h!]
\hspace{-3cm}
\begin{center}
\begin{tabular}{|c|c|c|c|c|}
\hline
$c_n$ & Asymptotic($n_l=0$) &Exact($n_l=0$) & Asymptotic($n_l=3$) &Exact($n_l=3$)                          \\\hline
$c_0$& $0.79(5)$ &  $0.734393$ & $0.64(12)$& 0.686724\\\hline
$c_1$ &$ 2.35(9) $ & $ 2.37374  $ & $  1.60(17) $& 1.71232 \\\hline
$c_2$ & $11.08(44)$ & $11.075 $ & $ 6.24(61) $ & 6.23808 \\\hline
$c_3$ & $71.6(2.8) $ & $                $ & $ 33.3(3.5)$ & $$ \\ \hline
$c_4$ & $ 589(23) $ & $ $ & $ 226(26) $ &\\\hline
$c_5$& $ 587(23)\times10$ & $  $ & $ 185(23) \times10 $ &\\ \hline
$c_6$& $ 688(27)\times10^2 $ & $  $ & $ 178(24) \times10^2  $ &\\ \hline
\end{tabular}
\end{center}
\caption{\it Renormalon-based predictions of the perturbative coefficients 
$c_n(m_{\PV})$ for $n_l=0$ and $n_l=3$.}
\label{tablecn}
\end{table}

\subsection{Hyperasymptotic expansion of $c_{\rm PV}$}
\label{PVsec}
As we have already mentioned, the divergent behavior of the perturbative series is regulated using the PV prescription for the Borel sum. The exact expression of the Borel transform is unknown. Therefore, one has to use approximations. In this context, it is of paramount importance to have a parametric control on the error with exponential accuracy. Consequently, we apply the hyperasymptotic approach developed in \cite{Ayala:2019uaw,Ayala:2019lak,Ayala:2019hkn}. $c_{\rm PV}$ will be computed truncating the hyperasymptotic expansion in a systematic way. The approximation used can be labeled by a pair of indices $(D,N)$: $c_{\rm PV} \longrightarrow c^{(D,N)}_{\rm PV}$. For the general definition and more details, one may look to \cite{Ayala:2019lak}.
In the case at hand, the Borel sum can be split into a partial sum truncated at the minimal term $(0,N_P)$, plus the leading terminant (1,0), plus a left-over of the perturbative expansion reaching the hyperasymptotic precision of $(1,N=N_{\rm max}-N_P)$ where $N_{\rm max}=2$ (the perturbative expansion is not known with high enough accuracy to go beyond that): 
\bea
\label{CBPVhyp}&&
c^{(1,N)}_{\PV}(m_{\PV})=c^{(0,N_P)}
\\
&&
\nn
+
\frac{1}{2}Z_{LS}\left[\frac{\als(m_{\PV})}{\als(\nu)}\right]^{\gamma_F}\Omega_+
+
\left(Z_m+\frac{1}{2}{\nu \over m_{\PV}}(Z_{\pi G}-2Z_{LS})\right)\frac{\Omega_m}{Z_m}
+
\frac{1}{2}Z_A\left[\frac{\als(m_{\PV})}{\als(\nu)}\right]^{-\gamma_F}\Omega_-
\\
&&
\nn
+
\sum_{n=N_P+1}^{N_P+N}\left(c_n-c_n^{(\rm asym)}\right)\als^{n+1}(\nu)
\,,
\eea
where 
\be
\label{NP}
N_P= \frac{2\pi}{\beta_0 \als(\nu)}\left(1-c\als(\nu)\right)
\,,
\ee
\be
c^{(0,N)}(\als)\equiv \sum_{n=0}^{N}c_n\als^{n+1}(\nu)
\,,
\ee
\be
\label{eq:Omegam}
\Omega_m \equiv Z_m\frac{\nu}{m_{\PV}} \frac{1}{\Gamma(1+b)}\left(\frac{\beta_{0}}{2 \pi}\right)^{N_P+1} \alpha_{X}^{N_P+2}(\nu) \int_{0, \mathrm{PV}}^{\infty} d x \frac{x^{b+N_P+1} e^{-x}}{1-x \frac{\beta_{0} \alpha_{X}(\nu)}{2 \pi }}
\,.
\ee
\be
\label{eq:Omegapm}
\Omega_{\pm} \equiv \frac{\nu}{m_{\PV}} \frac{1}{\Gamma(1+ b\pm \gamma_{\F})}\left(\frac{\beta_{0}}{2 \pi}\right)^{N_P+1} \alpha_{X}^{N_P+2}(\nu) \int_{0, \mathrm{PV}}^{\infty} d x \frac{x^{ b\pm \gamma_{\F}+N_P+1} e^{-x}}{1-x \frac{\beta_{0} \alpha_{X}(\nu)}{2 \pi }}
\,.
\ee

In the weak coupling limit, these expressions can be approximated by
\be
\label{eq:OmegamWeak}
\Omega_m
\simeq K_{m,\MS}^{\PV}(c)\left[\als(\nu)\right]^{\frac{1}{2}}\frac{\Lambda_{\MS}}{m_{\PV}}
\simeq K_{m,\MS}^{\PV}(c) \left[\als(\nu)\right]^{\frac{1}{2}-b}
\left(\frac{\beta_0}{4\pi}\right)^{-b}e^{-\frac{2\pi}{\beta_0\als(\nu)}}
\frac{\nu}{m_{\PV}}
\ee
where
\be
\label{KBPV}
K_{m,\MS}^{\PV}(c)=-\frac{Z_m}{\Gamma(1+b)}
\left(\frac{2\pi}{\beta_0}\right)2^{-b}\sqrt{\beta_0}\left[-\eta_c+\frac{1}{3}\right]
\,,
\ee
\be
\eta_c=-b+\frac{2\pi}{\beta_0}c-1
\,,
\ee
and
\be
\label{eq:OmegapmWeak}
\Omega_{\pm}
\simeq K_{\pm,\MS}^{\PV}(c)\left[\als(\nu)\right]^{\frac{1}{2}\mp \gamma_{\F}}\frac{\Lambda_{\MS}}{m_{\PV}}
\simeq K_{\pm,\MS}^{\PV}(c) \left[\als(\nu)\right]^{\frac{1}{2}-b \mp \gamma_{\F}}
\left(\frac{\beta_0}{4\pi}\right)^{-b}e^{-\frac{2\pi}{\beta_0\als(\nu)}}
\frac{\nu}{m_{\PV}}
\ee
where
\be
\label{KBPVpm}
K_{\pm,\MS}^{\PV}(c)=-\frac{1}{\Gamma(1+b\pm \gamma_{\F})}
\left(\frac{2\pi}{\beta_0}\right)^{\pm \gamma_{\F}+1}2^{-b}\sqrt{\beta_0}\left[-\eta^{\pm}_c+\frac{1}{3}\right]
\,,
\ee
\be
\eta^{\pm}_c=-b+\frac{2\pi}{\beta_0}c\mp \gamma_{\F}-1
\,.
\ee

 Eq. (\ref{CBPVhyp}) yields the maximal precision available at present. If we only use $c^{(0,N)}$ we are just working in standard perturbation theory. We can do so till we reach the asymptotic regime. We then have to truncate the perturbative series at $N_P$ (see Eq. (\ref{NP})): $c^{(0,N_P)}$. This is what it would be called superasymptotic approximation. Adding the terminants (the second line in Eq.  (\ref{CBPVhyp})) one reaches the hyperasymptotic precision (1,0). Note that in this paper, we face the case of having different renormalons located at the same position in the Borel plane, which can only be distinguished from the anomalous dimension. $\Omega_+$ is the dominant one, but the suppression of the other two renormalon structures is tiny, of order $\alpha^{\gamma_F}(m_{\PV})$ and $\alpha^{2\gamma_F}(m_{\PV})$. Such suppression factors can be identified in the prefactors multiplying $\Omega_m$ and $\Omega_\pm$ in Eq. (\ref{CBPVhyp}). They resum the large $\ln m_{\PV}$ logarithms associated with the anomalous dimension. 

For numerics, we will use Eqs. (\ref{eq:Omegam}) and (\ref{eq:Omegapm}). The difference with the weak coupling expressions in Eqs. (\ref{eq:OmegamWeak}) and (\ref{eq:OmegapmWeak}) are small. Moreover, the former gives a more precise result if the value of $c$ that appears in Eq. (\ref{eq:OmegamWeak}) is large (around 2 to 3), as it may happen in some cases in our computations. 


\section{Phenomenology}

\subsection{Determination of $c^{(b)}_{\PV}$, $c^{(c)}_{\PV}$ and $\hat c^{(b)}_{\PV}/\hat c^{(c)}_{\PV}$}

The key quantities that we want to compute in this section are $c^{(b)}_{\PV}$, $c^{(c)}_{\PV}$ and $\hat c^{(b)}_{\PV}/\hat c^{(c)}_{\PV}$. They enter into a variety of observables. In this paper we focus on the hyperfine splitting with the aim of determining some power suppressed nonperturbative constant and, above all, $\hat \mu^2_{G,\PV}$. We do so in the next section. 

The value of $\alpha_s(M_z)=0.1180(9)$ is taken from \cite{ParticleDataGroup:2020ssz}. After running down to scales of the order of 1 GeV using \cite{Herren:2017osy}, yields 
\be
\label{eq:LMS}
\Lambda_{\MS}=335^{+14}_{-13}\; {\rm MeV}.
\ee

For the bottom and charm masses we use the $\MS$ values obtained in Ref. \cite{Peset:2018ria} from the heavy quarkonium spectrum, as nonperturbative effects are suppressed for these observables. They read
\be
\label{eq:mbar}
\bar m_{b}=4.186(37) \;{\rm GeV}\,,  \quad {\rm and} \quad \bar m_{c}=1.223(33)\; ´{\rm GeV}
\,.
\ee
We then compute the PV masses using the normalization of the pole mass renormalon obtained in Ref. \cite{Ayala:2025trr}. We obtain 
\be
\label{eq:mPV}
m_{b,\PV}=4.843(41) \;{\rm GeV}\,,  \quad {\rm and} \quad  m_{c,\PV}=1.427(42)\; {\rm GeV}.
\ee
In these numbers, we have only included the error associated to the $\MS$ heavy quark masses, as it is the dominant source of error.

\begin{figure}
\begin{center}
\includegraphics[width=.7\textwidth]{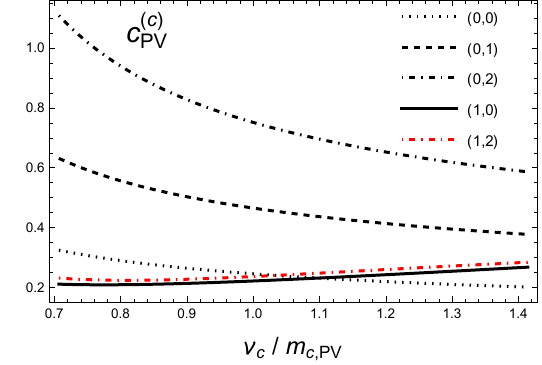}
\includegraphics[width=.7\textwidth]{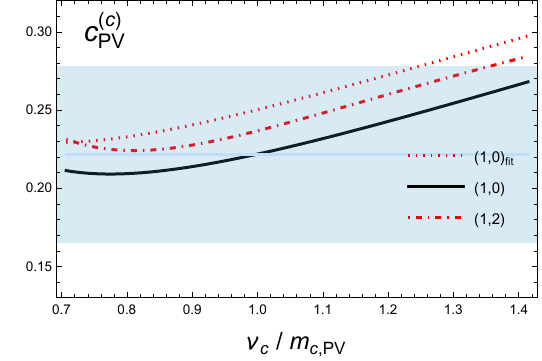}
\end{center}
\caption{{\bf Charm quark case.} \\
{\bf Upper figure}: Determinations of $c^{(c)}_{\PV}$ at ${\cal O}(\als)$ (dotted-black line), ${\cal O}(\als^2)$ (dashed-black line), ${\cal O}(\al^3)$ (dash-dotted black line), at order (1,0) in the hyperasymptotic expansion counting (continuous-black line), and at order (1,$N=2$)  in the hyperasymptotic expansion counting (dash-dotted red line). \\
{\bf Lower figure}: Zoom of the previous figure adding the determination of  $c^{(c)}_{\PV}$ at order (1,0) in the hyperasymptotic expansion counting (dotted-red line) using the normalizations of renormalons obtained in Eq. (\ref{fitnf3}). The horizontal line and blue band are our final value and error.}
\label{Fig:cPVcnf3}
\end{figure}

 \begin{figure}
\begin{center}
\includegraphics[width=.7\textwidth]{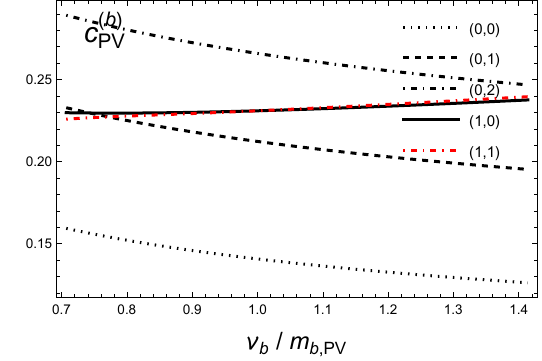}
\includegraphics[width=.7\textwidth]{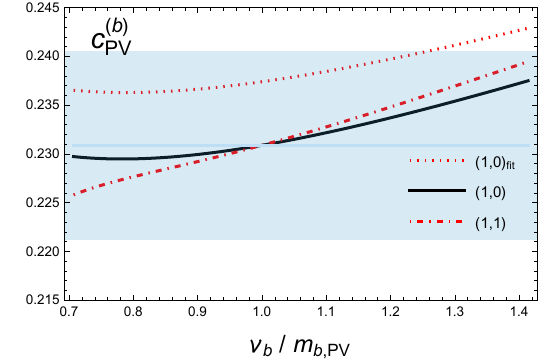}
\end{center}
\caption{ {\bf Bottom quark case.} \\
{\bf Upper figure}: Determinations of $c^{(b)}_{\PV}$ at ${\cal O}(\als)$ (dotted-black line), ${\cal O}(\als^2)$ (dashed-black line), ${\cal O}(\al^3)$ (dash-dotted black line), at order (1,0) in the hyperasymptotic expansion counting (continuous-black line), and at order (1,$N=1$)  in the hyperasymptotic expansion counting (dash-dotted red line).\\
{\bf Lower figure}: Zoom of the previous figure adding the determination of  $c^{(b)}_{\PV}$ at order (1,0) in the hyperasymptotic expansion counting (dotted-red line) using the normalizations of renormalons obtained in Eq. (\ref{fitnf3}). The horizontal line and blue band are our final value and error.}
\label{Fig:cPVbnf3}
\end{figure}

In Figs. \ref{Fig:cPVcnf3} and \ref{Fig:cPVbnf3}, we show the behavior of the hyperasymptotic expansion for the charm and bottom cases, respectively. We first explore the convergence of the perturbative expansion. For the case of the charm quark, we have bad convergence for the perturbative expansion. Whereas the ${\cal O}(\als)$ contribution (black-dotted line) is significantly smaller than one, the next two orders (dashed and dot-dashed black lines) are of similar size to the ${\cal O}(\als)$ term. This indicates that $N_P=0$ is a proper choice. This idea is reinforced after introducing the leading terminant, proportional to $Z_{LS}$, which is closer to the $N_P=0$ line albeit much more stable under scale variation. We take this result as our central value for $\nu=m_{\PV}$. We now turn to the discussion of how robust this result is, and to determine the error. 

The theoretical error is associated with our incomplete knowledge of the hyperasymptotic expansion. In our case, this reflects in the error of our determination of the renormalon normalization, our incomplete knowledge of subleading renormalons, and of the subleading terms of the hyperasymptotic expansion. These errors are highly correlated. We estimate them in a similar way to the way we computed the error of $Z_{LS}$ and the asymptotic coefficients. We consider the variations of $Z_{LS}$ due to its associated error from items (i), (ii) and (iii) in Sec. \ref{Sec:Norm}. The latter can be considered a measure of the effect of the subleading renormalons since the terminant is now determined using the second line of Eq. (\ref{CBPVhyp}) using the values of the normalization of the renormalons coming from the fit to the exact expression of the coefficient for $n=2$ listed in Eq. (\ref{fitnf3}) (the value of $Z_{LS}$ also changes accordingly). We then consider the impact of incorporating the last line of Eq. (\ref{CBPVhyp}). On top of that, these quantities should be scheme and scale independent. We estimate the scale dependence of the hyperasymptotic truncated expression by changing $\nu$ in the range $(m_{\PV}/\sqrt{2}, m_{\PV}\sqrt{2})$. The analysis for the bottom goes in a parallel way with the qualification that $N_P=1$ in this case. We disclose the error budget in Table \ref{tablecPV}. We observe that, in the case of the charm quark, the error is dominated by the scale dependence, whereas in the case of the bottom the error associated to the scale variation is of the same size as the error of considering subleading renormalons using item (iii) of Sec. \ref{Sec:Norm}. The value of $N_P$ that we use is relatively low compared to the natural values that follow from Eq. (\ref{NP}). On the other hand, it is the natural value for $N_P$ from the numerical analysis that follows from the magnitude of the different terms of the perturbative expansion shown in Figs. \ref{Fig:cPVcnf3} and \ref{Fig:cPVbnf3}. We have recomputed $c_{\PV}^{(1,0)}$ using different values of $N_P$ and the results are comfortably inside the error band. Indeed, in some cases better results are obtained. For instance, taking $N_P=1$ for the charm quark case, the scale dependence greatly diminishes.

The final theoretical error is obtained combining in quadrature the errors listed in Table \ref{tablecPV}. We observe that $c^{(Q)}(m_{Q,\PV})$ can be obtained with very high precision, particularly for the bottom quark case. It is also worth mention that $c^{(c)}(m_{c,\PV}) \simeq c^{(b)}(m_{b,\PV})$.

\begin{table}[htb!]
\begin{center}
$$
\begin{array}{|c|c|c|c|c|c|c|}
\hline
 \text{Q} & c_{\PV}^{(Q)}  & 
 \left\{
 Z_{LS}^{\left(x=1/\sqrt{2}\right)},
 Z_{LS}^{\left(x=\sqrt{2}\right)}
 \right\} 
 & Z_{LS}^{(\text{nlo})} &  (1,0)_\text{fit}& (1,N)&
 \frac{\nu}{m_{\PV}}=
   \left(\frac{1}{\sqrt{2}},\sqrt{2}\right) \\
   \hline
 b & 0.2309(95)  & \{-0.0012,0.0007\} & 0.0014& 0.0065  & 0.  & \{-0.0012,0.0066\} \\
 \hline
 c & 0.222(56)  & \{0.001,-0.001\} & -0.0018 & 0.028  & 0.015 & \{-0.011,0.046\}  \\
 \hline
\end{array}
$$
\end{center}
\caption{\it Theoretical error budget of $c^{(Q)}_{PV}$ for the botttom and charm quark case. In the first column we give our final prediction for $c^{(Q)}_{PV}$ and its theoretical error. In the following columns we disclose the error associated to each different source. In the previous to last column $N=1$ for bottom and $N=2$ for charm.}
\label{tablecPV}
\end{table}
Besides the pure theoretical error studied above, we also have some errors associated with our incomplete knowledge of the strong coupling constant and of the heavy quark masses. We list them in Table \ref{tablecPVexp}. We observe that they are small compared to the theoretical error. 
\begin{table}[htb!]
\begin{center}
$$
\begin{array}{|c|c|c|c|}
\hline
 \text{Q} & c_{\PV}^{(Q)}  & \Lambda_{QCD} &
   m_{Q,\PV} \\
   \hline
 b & 0.2309  & 0.002 & ^{-0.0001}_{+0.0007}\\
 \hline
 c & 0.222   & 0.005 & ^{+0.0033}_{-0.0038} \\
 \hline
\end{array}
$$
\end{center}
\caption{\it Error budget of $c^{(Q)}_{PV}$ for the botttom and charm quark case associated with $\Lambda_{\rm QCD}$ and the heavy quark masses. }
\label{tablecPVexp}
\end{table}

For the final error, we combine in quadrature the total theoretical error listed in Table \ref{tablecPV} with the error associated with $\Lambda_{\rm QCD}$ and the heavy quark masses listed in Table \ref{tablecPVexp}. The final result reads
\be
c_{\PV}^{(b)}(m_{b,\PV})=0.2309(97) \,, \qquad  c_{\PV}^{(c)}(m_{c,\PV})=0.222(57) 
\,.
\ee
It is remarkable that these coefficients can be obtained with a very high precision. Indeed, for $c_{\PV}^{(b)}$, a 5 per mille precision is achieved. 

\medskip

\begin{figure}
\begin{center}
\includegraphics[width=.7\textwidth]{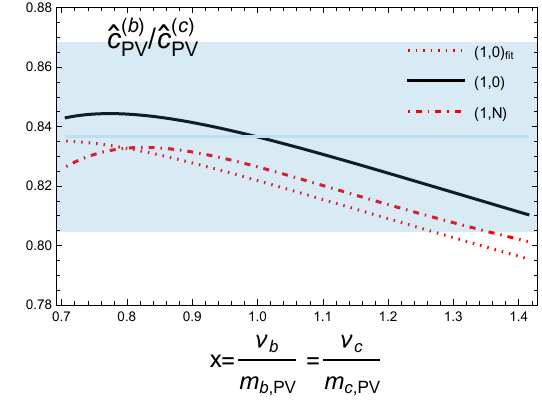}
\end{center}
\caption{ 
Determination of $\hat c^{(b)}_{\PV}/\hat c^{(c)}_{\PV}$ at order (1,0) in the hyperasymptotic expansion counting (continuous black line), at order (1,$N$) (where $N=1$ for $\hat c^{(b)}_{\PV}$ and $N=2$ for $\hat c^{(c)}_{\PV}$) in the hyperasymptotic expansion counting (dash-dotted red line), and at order (1,0) in the hyperasymptotic expansion counting (dotted-red line) using the normalization of the renormalons obtained in Eq. \ref{fitnf3}. The horizontal line and blue band are our final value and error.
}
\label{Fig:hatcPVbcnf3}
\end{figure}

Of great interest to us is the ratio $\hat c^{(b)}_{\PV}/\hat c^{(c)}_{\PV}$. We plot it in Fig. \ref{Fig:hatcPVbcnf3}. The determination of its error goes parallel to the analysis made for $c^{(Q)}_{\PV}$. The theoretical error is disclosed in Table \ref{tablehatcPVratio}. Note that in this analysis, the scale variation of $\hat c^{(b)}_{\PV}$ and $\hat c^{(c)}_{\PV}$ 
is made in a correlated way (if one renormalization scale is better or worse than another, it should be so for both quarks, as they share the same physics content). In Table \ref{tablehatcPVratioexp}, we list the errors associated to $\Lambda_{\rm QCD}$ and the heavy quark masses. 

\begin{table}[htb!]
\begin{center}
$$
\begin{array}{|c|c|c|c|c|c|c|}
\hline
 \hat c^{(b)}_{\PV}/\hat c^{(c)}_{\PV}  & 
 \left\{
 Z_{LS}^{\left(x=1/\sqrt{2}\right)},
 Z_{LS}^{\left(x=\sqrt{2}\right)}
 \right\} 
 & Z_{LS}^{(\text{nlo})} & (1,0)_\text{fit}& (1,N)&
 \frac{\nu}{m_{\PV}}=
   \left(\frac{1}{\sqrt{2}},\sqrt{2}\right) 
   \\
   \hline
   0.836(32) & \{-0.002, 0.001\} & 0.002 & -0.015 & -0.010 & \{0.006,-0.026\}   \\
 \hline
\end{array}
$$
\end{center}
\caption{\it Theoretical error budget of $\hat c^{(b)}_{\PV}/\hat c^{(c)}_{\PV}$. In the first column we give our final prediction for 
$\hat c^{(b)}_{PV}/\hat c^{(c)}_{PV}$ and its theoretical error. In the following columns we disclose the error associated to each different source. In the previous to last column $N=1$ for $\hat c^{(b)}_{\PV}$ and $N=2$ for $\hat c^{(c)}_{\PV}$. }
\label{tablehatcPVratio}
\end{table}
\begin{table}[htb!]
\begin{center}
$$
\begin{array}{|c|c|c|c|c|}
\hline
  \hat c^{(b)}_{\PV}/\hat c^{(c)}_{\PV}   & \Lambda_{QCD} &
   m_{b,\PV} & m_{c,\PV}
   \\
   \hline
   0.836 &  0.001 & ^{-0.0009}_{+0.0013} & ^{+0.003}_{-0.003}  \\
 \hline
\end{array}
$$
\end{center}
\caption{\it 
Error budget of $\hat c^{(b)}_{\PV}/\hat c^{(c)}_{\PV}$ associated with $\Lambda_{\rm QCD}$ and the heavy quark masses.}
\label{tablehatcPVratioexp}
\end{table}

For the final error, we combine in quadrature the total theoretical error listed in Table \ref{tablehatcPVratio} with the error associated with $\Lambda_{\rm QCD}$ and the heavy quark masses listed in Table \ref{tablehatcPVratioexp}. The final result reads
\be
\label{eq:cRatioFinal}
\hat c^{(b)}_{\PV}/\hat c^{(c)}_{\PV}=0.836(32)
\,.
\ee
Note that the error is completely dominated by theory, yet the number we obtain is quite precise.

\subsection{Determination of $\hat \mu^2_{G,\PV}$}

Besides its interest for $B/D$ meson spectroscopy, $\mu^2_{G,\PV}$ is important for precise determinations of $V_{cb}$ from semileptonic decays. This venue will be explored in the future. 

The main aim of this section is to determine $\hat \mu^2_{G,\PV}$ with power-like accuracy. To do so, we need $c_{\PV}$ with power-like accuracy, as well as the $1/m^2$ terms that appear in Eq. (\ref{HFPV}). We then compare our theoretical predictions with the
experimental data using the values of the $B/D$ meson masses quoted in Ref. \cite{ParticleDataGroup:2020ssz}.
In comparing theory and experiment, we will let $\hat \mu_{G,\PV}^{2}$ to be a free parameter, which will then be fitted to the experimental data. 

The strategy we follow is that we first determine (a combination of) the $1/m^2$ terms that appear in Eq. (\ref{HFPV}) using the following equality:
\be
\frac{M_B^{*2}-M_B^2}{M_D^{*2}-M_D^2}
=
\frac{\hat c_{\PV} (m_{b,\PV})}{ \hat c_{\PV}(m_{c,\PV})}
\left[
1+
\left(
\frac{1}{m_{b,\PV}}-\frac{1}{m_{c,\PV}}
\right)
A
\right]
\ee
where 
\be
A \equiv 
\frac{\rho_{\pi G,\PV}^3-2\rho^3_{LS,\PV}+c_{F,\PV}\rho_{A,\PV}^3+c^{-1}_{F,\PV}\rho^3_{LS,\PV}}{2\mu^2_{G,\PV}}-\bar \Lambda_{\PV}
\,.
\ee
In this equation, we have approximated $\hat c_{\PV} (m_{b,\PV})\simeq \hat c_{\PV}(m_{c,\PV})$ in $A$. Since we have that $\hat c_{\PV} (m_{b,\PV})/ \hat c_{\PV}(m_{c,\PV}) \simeq 0.84$, the error associated to this approximation is small in comparison with other sources of errors.

The experimental error associated to the masses of the $B/D$ mesons of this determination is negligible. The error of $A$ is completely correlated with the error of $\hat c_{\PV} (m_{b,\PV})/ \hat c_{\PV}(m_{c,\PV})$ obtained in Eq. (\ref{eq:cRatioFinal}).  Therefore, we obtain 
\be
\label{eq:A}
A= -0.121(85)
\,.
\ee
One may wonder about adding an error associated with $1/m^2$ effects. This is already accounted for in the final error of $\hat c_{\PV} (m_{b,\PV})/ \hat c_{\PV}(m_{c,\PV})$. In any case one would expect corrections of order $m_c/m_b \sim 1/3$ which is way smaller than the total error quoted in Eq. (\ref{eq:A}). Note that $\hat c_{\PV} (m_{b,\PV})/ \hat c_{\PV}(m_{c,\PV})$ is known with power-like accuracy, and we need power-like accuracy in the perturbative expression to determine a power suppressed nonperturbative correction. 

$\bar \Lambda_{\PV}$ can also be obtained directly from $B$ meson mass. One obtains
\be
\bar \Lambda_{\PV}=470(62)\; {\rm MeV}
\,,
\ee
where we have combined in quadrature the error we have for $m_{b,\PV}$ in Eq. (\ref{eq:mPV}) and the error associated to possible ${\cal O}(1/m_{b,\PV})$ effects. The latter were assigned to be of around $\sim 46$ MeV in Ref. \cite{Ayala:2019hkn}. 

We can then also determine $A+\bar \Lambda_{\PV}$. We obtain
\be
\frac{\rho_{\pi G,\PV}^3-2\rho^3_{LS,\PV}+c_{F,\PV}\rho_{A,\PV}^3+c^{-1}_{F,\PV}\rho^3_{LS,\PV}}{2\mu^2_{G,\PV}}
=0.35(11)
\,,
\ee
where we have combined in quadrature the error of $A$ and the error of $\bar \Lambda$.

We are now in the position to determine $\hat \mu^2_{G,\PV}$. Using (we take the bottom quark case because the nonperturbative corrections are more suppressed)
\be
M_B^{*2}-M_B^2
=
\frac{4}{3}
\hat c_{\PV} (m_{b,\PV})\hat \mu^2_{G,\PV}
\left[
1+
\frac{1}{m_{b,\PV}}
A
\right]
\,,
\ee
we obtain
\be
\label{eq:muGPVFinal}
\mu_{G,\PV}^2=0.507(7)
\,.
\ee 
We emphasize that this number is independent of the heavy quark mass. It should be proportional to $\Lambda_{\rm QCD}^2$ and of natural size. It should also be independent of the scheme/scale used for the strong coupling. The error of this determination comes from the error of $\hat c_{\PV} (m_{b,\PV})$ and $A$. They are correlated. Therefore, we have taken into account this correlation by making the error variations simultaneously in a similar manner as we have done the variation of $\hat c_{\PV} (m_{b,\PV})$ and $A$. The disclosure of the theoretical error can be found in Table \ref{tablemuG2}. The error associated to $\Lambda_{\rm QCD}$ and the heavy quark masses can be found in Table \ref{tablemuG2exp}. The final error is obtained by combining them in quadrature. 

\begin{table}[htb!]
\begin{center}
$$
\begin{array}{|c|c|c|c|c|c|}
\hline
 \hat \mu_{G,\PV}^2  & 
 \left\{
 Z_{LS}^{\left(x=1/\sqrt{2}\right)},
 Z_{LS}^{\left(x=\sqrt{2}\right)}
 \right\} 
 & Z_{LS}^{(\text{nlo})} & (1,0)_\text{fit}& (1,N)&
 \frac{\nu}{m_{\PV}}=
   \left(\frac{1}{\sqrt{2}},\sqrt{2}\right) 
   \\
   \hline
   0.507(6) & \{0.001, -0.001\} & -0.001 & 0.002 & 0.003 & \{-0.002,0.005\}   \\
 \hline
\end{array}
$$
\end{center}
\caption{\it Theoretical error budget of $\hat \mu_{G,\PV}^2$. In the first column we give our final prediction for 
$A$ and its theoretical error. The following columns disclose the error associated to each different source. }
\label{tablemuG2}
\end{table}

\begin{table}[htb!]
\begin{center}
$$
\begin{array}{|c|c|c|c|}
\hline
 \hat \mu_{G,\PV}^2  & \Lambda_{QCD} &
   m_{b,\PV} & m_{c,\PV}
   \\
   \hline
   0.507 &  0.003 & ^{+0.0004}_{-0.0006} & 0.0005  \\
 \hline
\end{array}
$$
\end{center}
\caption{\it Error budget of $\hat \mu_{G,\PV}^2$ associated with $\Lambda_{\rm QCD}$ and the heavy quark masses.}
\label{tablemuG2exp}
\end{table}

\section{Conclusions}

In this paper, we have obtained the first determination of $Z_{LS}$, the normalization of the leading infrared renormalon of the chromomagnetic moment of a heavy quark. We obtain 
\be
Z_{LS}(n_l=0)= 1.58(19) \,, \qquad 
Z_{LS}(n_l=3)= 1.29(44)
\,.
\ee
For the other infrared renormalons located at the same position in the Borel plane but with a weaker singularity, we have also given some rough estimates. 

Using these results and the knowledge of the renormalon structure of $c_F$, we have given some estimates of higher order terms of the perturbative series of $c(m)$. They can be found in Table \ref{tablecn}.

We have then determined $\hat c_{F,\PV}$ with superasymptotic (truncating the perturbative expansion at the minimal term) and hyperasymptotic accuracy (by also including the leading terminant), i.e. with power-like accuracy. See Eq. (\ref{CBPVhyp}). This allows us to obtain from experiment power-like suppressed nonperturbative parameters. We have obtained
\be
A=-0.121(85)
\,,
\ee
and
\be
\hat \mu^2_{G,\PV}=0.507(7) \; {\rm GeV}^2
.
\ee 
We emphasize that these results are independent of the renormalization scale and scheme used for the strong coupling.
They are of interest for a variety of $B$ and $D$ physics observables. The application of our numbers to such observables will be considered elsewhere. 

Determinations of $\hat \mu^2_{G}(m_b)$ have been given in the past. See, for instance, Refs. \cite{Gambino:2017vkx,FermilabLattice:2018est,Hayashi:2022hjk}. Nevertheless, none of them incorporate subleading renormalons. Therefore, their numbers suffer from an intrinsic summation scheme dependence that limits the accuracy they can reach. This reflects in the fact that the numbers given in those references quote errors much larger than those we obtain in this paper. A number for the coefficient $A$ has been obtained in Ref. \cite{Ayala:2025yjy} using a dispersive approach for the strong coupling constant and including the effect of the leading renormalons. The number obtained is between 1-2 sigmas above our figure with comparable errors. This dispersive approach requires a model-dependent parameterization of the strong coupling constant. The associated error should be small, yet the dependence on the parameter $M_1$ is relatively large compared with the expected power suppression of these effects, producing the bulk of the error in that reference. On the other in Ref. \cite{Ayala:2025yjy} there is no a specific error associated with the incomplete knowledge of the perturbative expansion and the operator product expansion. This should give the largest contribution to the error according to the hyperasymptotic expansion analysis. 

Our results improve over the analysis made in Ref.\cite{Grozin:1997ih}, which indeed can be understood as a computation made with superasymptotic precision. Here we formalize these results and go beyond them by including the leading terminant. The introduction of the terminant is crucial to make the result more stable under scale variations. The introduction of the terminant allows us to have power-like precision in the determination of nonperturbative constants. One advantage of the present method is that it minimizes the dependence on the normalization of the renormalon. Another major advantage is that it allows to obtain the ``real'' size of the nonperturbative corrections, in the sense that the nonperturbative correction scales as powers of $\Lambda_{QCD}$. This makes it clearer what the size of the nonperturbative corrections is without contamination from perturbation theory.
  
The theoretical error of $\hat \mu_{G,\PV}^2$ and $A$ is dominated by theory. The experimental error is negligible. Therefore, all these theoretical errors would benefit from higher order computations. To know the anomalous dimension of the nonperturbative constants to a higher order would also be of help. These computations would improve the determination of the normalization of the leading renormalon as well as, hopefully, yield a better signal for the other renormalons.

\medskip
   
\noindent
{\bf Acknowledgments.}\\
This work was supported in part by the Spanish Ministry of Science and Innovation PID2023-146142NB-I00 and by FONDECYT (Chile) Grant No.~1240329.

\appendix


\end{document}